\RequirePackage{ifpdf}
\ifpdf 
\documentclass[pdftex]{sigma}
\else
\documentclass{sigma}
\fi

\newcommand{\al}{\alpha}
\newcommand{\be}{\beta}

\newcommand{\ep}{\epsilon}

\newcommand{\la}{\lambda}
\newcommand{\om}{\omega}
\newcommand{\si}{\sigma}

\newcommand{\vp}{\varphi}
\newcommand{\ze}{\zeta}
\newcommand{\La}{\Lambda}
\newcommand{\Si}{\Sigma}
\newcommand{\bx}{{\boldsymbol{x}}}

\newcommand{\bv}{{\boldsymbol{v}}}
\newcommand{\bz}{{\boldsymbol{z}}}

\newcommand\Th{\tilde h}
\newcommand\CC{\mathbb C}
\newcommand\NN{\mathbb N}
\newcommand\RR{\mathbb R}

\newcommand{\cB}{{\mathcal B}}

\newcommand{\cH}{{\mathcal H}}

\newcommand{\cM}{{\mathcal M}}

\newcommand{\cP}{{\mathcal P}}

\newcommand{\cS}{{\mathcal S}}
\newcommand{\cT}{{\mathcal T}}

\newcommand\cL{\mathcal L}
\newcommand\cX{\mathcal X}
\newcommand{\fW}{{\mathfrak W}}

\def\BH{\,\overline{\!H}{}}

\def\Bx{\overline x}
\def\BcH{\overline{\mathcal H}}
\newcommand{\pa}{\partial}
\newcommand\pd\partial

\def\ket#1{|#1\rangle}

\def\BStrut{\vrule height14pt depth6pt width0pt}

\newcommand{\ms}{\mspace{1mu}}
\newcommand\ran\rangle
\newcommand\lan\langle
\DeclareMathOperator\End{End}

\newcommand{\iu}{{\mathrm i}}
\newcommand\I{\mathrm i}
\newcommand{\e}{{\mathrm e}}
\newcommand\dd{\mathrm d}

\newcommand\SU{\mathrm{SU}}
\begin{document}
\allowdisplaybreaks

\renewcommand{\PaperNumber}{073}
\renewcommand{\thefootnote}{$\star$}

\FirstPageHeading

\ShortArticleName{QES $N$-Body Spin Hamiltonians with Short-Range Interaction Potentials}

\ArticleName{Quasi-Exactly Solvable $\boldsymbol{N}$-Body Spin Hamiltonians\\
with Short-Range Interaction Potentials\footnote{This paper is a contribution 
to the Proceedings of the Workshop on 
Geometric Aspects of Integ\-rable Systems
 (July 17--19, 2006, University of Coimbra, Portugal).
The full collection is available at 
\href{http://www.emis.de/journals/SIGMA/Coimbra2006.html}{http://www.emis.de/journals/SIGMA/Coimbra2006.html}}}

\Author{A. ENCISO, F. FINKEL, A. GONZ\'ALEZ-L\'OPEZ~$^*$ and M.A. RODR\'IGUEZ} 
\AuthorNameForHeading{A. Enciso, F. Finkel, A. Gonz{\'a}lez-L\'opez and M.A. Rodr\'\i guez}

\Address{Depto.~F\'\i sica Te\'orica II, Universidad Complutense, 28040 Madrid, Spain} 
\EmailMarked{\href{artemio@fis.ucm.es}{artemio@fis.ucm.es}} 

\ArticleDates{Received September 15, 2006, in f\/inal form October 23,
2006; Published online November 03, 2006}

\Abstract{We review some recent results on quasi-exactly solvable
spin models presenting near-neighbors interactions. These systems
can be understood as cyclic generalizations of the usual
Calogero--Sutherland models. A nontrivial modif\/ication of the
exchange operator formalism is used to obtain several inf\/inite
families of eigenfunctions of these models in closed form.}


\Classification{81Q05; 35Q40} 

\section{Introduction}

In the early 1970s, F.~Calogero~\cite{Ca71} and
B.~Sutherland~\cite{Su71} introduced the quantum integrable
systems that nowadays bear their names. Apart from their intrinsic
mathematical interest~\cite{Mo75,Hi87,BF97}, Calogero--Sutherland
(CS) models have played a central role in Physics due to their
relevant applications to as diverse topics as soliton
theory~\cite{Ka95,Po95}, quantum f\/ield and string
theory~\cite{GN94,DP98}, quantum Hall ef\/fect~\cite{AI94},
fractional statistics~\cite{Po89} and random matrix
theory~\cite{Sh01}. The f\/irst satisfactory explanation of the
integrability of these models was given by Olshanetsky and
Perelomov~\cite{OP83}, who connected these models with the radial
Laplacian of symmetric spaces associated to the root system $A_N$.\
This unif\/ied view enabled them to introduce important
generalizations, including dif\/ferent root systems and elliptic
potentials.

During the last decade CS models have been extended to the case of
particles with internal degrees of freedom, which we shall
henceforth call spin. There are two dif\/ferent approaches to spin
CS models, namely the supersymmetric~\cite{BTW98} and Dunkl
operator~\cite{Du89,Po92} formalisms. These methods have allowed
to solve totally or partially several rational, trigonometric and
elliptic spin models, both in their $A_N$ and $BC_N$
versions~\cite{DLM01,Ya95,FGGRZ01,FGGRZ01b}. The interest in spin
CS models also increased as a consequence of their direct
connection with the Haldane--Shastry (HS) spin
chain~\cite{Ha88,Sh88}, which was laid bare by Polychronakos
through the so called ``freezing trick''~\cite{Po93}. This
technique was also used in the construction of solvable spin
chains associated to dif\/ferent potentials and root
systems~\cite{Po94,EFGR05}.

Auberson, Jain and Khare~\cite{JK99,AJK01} introduced partially
solvable versions of the CS models in which each particle only
interacts with its nearest and next-to-nearest neighbors. Similar
scalar models were also studied by Ezung, Gurappa, Khare and
Panigrahi~\cite{EGKP05}. There are two reasons that make these
kind of systems very promising from a physical point of view.
First, some of them are related to the short-range Dyson model
in random matrix theory~\cite{BGS99}. Second, the HS
chains associated to these models would occupy an interesting
intermediate position between the Heisenberg chain (short-range,
position-independent interactions) and the usual HS chains
(long-range, position-dependent interactions). A f\/irst step
towards the construction of these chains was Deguchi and Ghosh's
def\/inition of spin 1/2 versions of the Jain and Khare
Hamiltonians~\cite{DG01} using the supersymmetric formalism.
Unfortunately, all these authors solely managed to construct a few
exact solutions, and all of them with trivial spin dependence, and
the procedures developed to obtain exact solutions are by no means
systematic.

In \cite{EFGR05b,EFGR06} we introduced three new families of
spin near-neighbors models and used a nontrivial modif\/ication of
the Dunkl operator method to obtain a wide range of fully explicit
solutions. In this article we aim to review the main ideas
underlying our constructions and explain their connection with the
usual CS models. In Section~\ref{S:main} we def\/ine the
Hamiltonians that we shall deal with and state our main result,
which is a description of the algebraic states of the models. In
Section~\ref{S:Calogero} we review the philosophy underlying the
calculation of exact solutions of Schr\"odinger's equation by
algebraic methods. In Section~\ref{S:Inv} we sketch the main
logical steps that the construction of the invariant f\/lags rests
upon. We f\/inish the paper by showing in Section~\ref{S:spec} how
the actual computation of the algebraic eigenfunctions can be
carried out. For the sake of brevity, complete proofs are not
given in this review.

\section{Main result}\label{S:main}

Let $\Si$ be the Hilbert space of internal degrees of freedom of
$N$ particles of spin $M\in\frac12\NN$. Let us f\/ix a basis
\[
\cB=\big\{\ket{s_1,\dots,s_N}:s_i\in\{-M,-M+1,\dots,M\}\big\}
\]
of $\Si$ and def\/ine the spin exchange operators $S_{ij}$ as
\[
S_{ij}\ket{s_1,\dots,s_i,\dots,s_j,\dots,s_N}=\ket{s_1,\dots,s_j,\dots,s_i,\dots,s_N} .
\]
These operators can be expressed in terms of the (normalized)
fundamental generators of $\SU(2M+1)$ as
$S_{ij}=(2M+1)^{-1}+\sum_aJ^a_iJ^a_j$, the index sum ranging from
1 to $4M(M+1)$.

The Hamiltonians of the models we shall be concerned
with are given by (cf.~\cite{EFGR06})
\begin{equation}\label{Hep}
H_\ep=-\sum_i\pa_{x_i}^2+V_\ep ,\qquad\ep=0,1,2,
\end{equation}
where
\begin{subequations}\label{Vs}
\begin{gather}
V_0=\omega^2
r^2+\sum_i\frac{2a^2}{(x_i-x_{i-1})(x_i-x_{i+1})}+
\sum_i\frac{2a}{(x_i-x_{i+1})^2} (a-S_{i,i+1}),\label{V0}\\
V_1=\omega^2 r^2+\sum_i\frac{b(b-1)}{x_i^2}
+\sum_i\frac{8a^2x_i^2}{(x_i^2-x_{i-1}^2)(x_i^2-x_{i+1}^2)}\notag\\
\phantom{V_1=}{}+4a\sum_i\frac{x_i^2+x_{i+1}^2}{{(x_i^2-x_{i+1}^2)}^2}
 (a-S_{i,i+1}),\label{V1}\\
V_2=2a^2\sum_i\cot(x_i-x_{i-1})\cot(x_i-x_{i+1})
+2a\sum_i\csc^{2}(x_i-x_{i+1}) (a-S_{i,i+1}),\label{V2}
\end{gather}
\end{subequations}
with $r^2=\sum_i x_i^2$ and $a,b>1/2$. Here and in what follows,
all sums and products run from~$1$ to $N$ unless otherwise stated,
with the identif\/ications $x_0\equiv x_N$ and $x_{N+1}\equiv x_1$.
There is a~hyperbolic potential analogous to~\eqref{V2} that is
recovered substituting $x_i$ by $-x_i$ and $V_2$ by $-V_2$. The
\emph{scalar reductions} $H_\ep^{\mathrm{sc}}$ of these models are
obtained from the above Hamiltonians by the substitution
$S_{i,i+1}\to1$.

A few remarks on the conf\/iguration spaces of these models are now
in order. In all three models the potential diverges as
$(x_i-x_{i+1})^{-2}$ on the hyperplanes $x_i=x_{i+1}$, so that the
particles~$i$ and $i+1$ cannot overtake one another. Since we are
interested in models with nearest and next-to-nearest neighbors
interactions, we shall henceforth assume that $x_1<\cdots<x_N$.
For the second potential~\eqref{V1} we shall take in addition
$x_1>0$, due to the double pole at $x_i=0$.

A f\/irst observation concerning the eigenstates of the
models~\eqref{Hep} is that if $\psi$ is an eigenfunction of
$H_\ep^{\mathrm{sc}}$ with energy $E$, then the \emph{factorized
state} $\Psi=\psi\ket s$ is an eigenfunction of $H_\ep$ with the
same energy for any spin state $\ket s$ symmetric under
permutation of particles. A second observation is that $H_2$
commutes with the total momentum $P=-\I\sum_i\pd_{x_i}$. Hence the
movement of the center of mass decouples, and only the
eigenfunctions of $H_2$ with zero total momentum need to be
considered.

The next theorem summarizes the main results presented in~\cite{EFGR05b,EFGR06}, which yield a fully explicit
description of several families of algebraic eigenfunctions. It is
not dif\/f\/icult to realize, however, that these eigenfunctions do not
exhaust the whole spectrum of the models. We shall denote by
$\Lambda$ the projection operator on states totally symmetric under
the simultaneous permutation of both the spatial and spin
coordinates. We def\/ine the spin vectors $\ket{s_i}$, $\ket{s_{ij}}$
as
\[
\La(x_1\ket s)=\sum_ix_i\ket{s_i} ,\qquad \La(x_1x_2\ket
s)=\sum_{i<j}x_ix_j\ket{s_{ij}} .
\]
We shall consider the subspace $\Si'\subset\Si$ of spin
vectors $\ket s$ such that $\sum_i\ket{s_i}$ is symmetric. 
A~thorough characterization of this space is given in~\cite{EFGR06}.
\begin{theorem}\label{main}
Let $l$, $m$ be nonnegative integers and let $\ket s$ be an arbitrary
spin vector, and denote by $\Bx$ the center of mass coordinate $\frac 1N\sum_ix_i$.
Then the following statements hold:

{\rm 1.} Let $\al=N(a+\frac12)-\frac32$,
$\be\equiv\be(m)=1-m-N(a+\frac12)$, $t=\frac{2r^2}{N\Bx^2}-1$, and
$\mu_{lm}=\e^{-\frac\om2r^2}\Bx^mL_l^{-\be}(\om
r^2)\prod_i|x_i-x_{i+1}|^a$. The Hamiltonian $H_0$ possesses the
following families of spin eigenfunctions with eigenvalue
$E_{lm}=E_0+2\om(2l+m)$, where $E_0=N\omega(2a+1)$ is the ground state energy:
\begin{gather*}
\Psi^{(0)}_{lm}=\mu_{lm}\ms
P^{(\al,\be)}_{[\frac m2]}(t) \La\ket s ,\qquad m\geq0,\\
\Psi^{(1)}_{lm}=\mu_{l,m-1}\bigg[P^{(\al+1,\be)}_{[\frac{m-1}2]}(t) \La(x_1\ket
s)
-\Bx P^{(\al+1,\be)}_{[\frac{m-1}2]}(t) \La\ket s\bigg] ,\qquad m\geq1 ,\\
\Psi^{(2)}_{lm}=\mu_{l,m-2}\bigg[P^{(\al+2,\be)}_{[\frac m2]-1}(t)\big(\La(x_1^2\ket s)-2\Bx \La(x_1\ket s)\big)\\
\phantom{\Psi^{(2)}_{lm}=}{}+\Bx^2\bigg(P^{(\al+2,\be)}_{[\frac{m}{2}]-1}(t)
-\frac{2(\al+1)}{2[\tfrac{m-1}2]+1} P^{(\al+1,\be)}_{[\frac{m}{2}]
-1}(t)\bigg)\La\ket s\bigg] ,\qquad m\geq2 ,\\
\Psi^{(3)}_{lm}=\mu_{l,m-3}\bigg[
\frac2{3N} P^{(\al+3,\be)}_{[\frac{m-3}2]}(t)\sum_i
x_i^3+\Bx^3\vp_m(t)\bigg] \La\ket s , \qquad m\geq3 ,\\
\Psi^{(4)}_{lm}=\mu_{l,m-4}\bigg[\frac3{2([\frac{m-3}2]+\frac12)} \Bx^2P^{(\al+3,\be)}_{[\frac
m2]-2}(t) \La(x_1^2\ket s)\\
\phantom{\Psi^{(4)}_{lm}=}{}+\Big(\frac32 \Bx^3\phi_m(t)-\frac{1}N P^{(\al+4,\be)}_{[\frac
m2]-2}(t)\sum_i x_i^3\Big)\La(x_1\ket s)\\
\phantom{\Psi^{(4)}_{lm}=}{}+\Big(\frac1N \Bx P^{(\al+4,\be)}_{[\frac
m2]-2}(t)\sum_ix_i^3+\frac32 \Bx^4\chi_m(t)\Big)\La\ket
s\bigg] ,\qquad m\geq4 .
\end{gather*}
When $\ket s\in\Si'$ there is an additional family of
eigenfunctions given by
\begin{gather*}
\widetilde\Psi^{(2)}_{lm}=\mu_{l,m-2}\bigg[P^{(\al+2,\be)}_{[\frac
m2]-1}(t)
\big(\La(x_1x_2\ket s)-2\Bx \La(x_1\ket s)\big)\\
\phantom{\widetilde\Psi^{(2)}_{lm}=}{}+\Bx^2\bigg(P^{(\al+2,\be)}_{[\frac{m}{2}]-1}(t)
+\frac{2(\al+1)}{\big(2[\tfrac{m-1}2]+1\big)(N-1)} P^{(\al+1,\be)}_{[
\frac m2]-1}(t)\bigg)\La\ket s\bigg] ,\qquad m\geq2 .
\end{gather*}
The functions $\vp_m$, $\phi_m$ and $\chi_m$ are polynomials given
explicitly by
\begin{gather*}
\varphi_m=\frac{m+2\al+2}{m-1} P^{(\al+2,\be-2)}_{\frac m2}
-P^{(\al+3,\be-1)}_{\frac
m2-1}-\frac{4\al+7}{m-1} P^{(\al+2,\be-1)}_{\frac m2-1}
+\frac13 P^{(\al+3,\be)}_{\frac m2-2} ,\\
\phi_m=P^{(\al+4,\be-1)}_{\frac m2-1}-2P^{(\al+3,\be-1)}_{\frac
m2-1}-\frac{m+2\al+3}{(m-1)(m-3)} P^{(\al+2,\be-1)}_{\frac
m2-1}\\
\phantom{\phi_m=}{}-\frac13 P^{(\al+4,\be)}_{\frac
m2-2}+\frac{m+2\al-1}{m-3} P^{(\al+3,\be)}_{\frac m2-2} ,\\
\chi_m=\frac{3m+2\al}{(m-1)(m-3)} P^{(\al+2,\be-1)}_{\frac
m2-1}+\frac{2m-7}{m-3} P^{(\al+3,\be-1)}_{\frac
m2-1}-P^{(\al+4,\be-1)}_{\frac m2-1}\\
\phantom{\chi_m=}{}-\frac{m+2\al+2}{(m-1)(m-3)} P^{(\al+2,\be)}_{\frac
m2-2}-\frac{m+2\al}{m-3} P^{(\al+3,\be)}_{\frac
m2-2}+\frac13 P^{(\al+4,\be)}_{\frac m2-2} ,
\end{gather*}
for even $m$, and
\begin{gather*}
\varphi_m=2P^{(\al+2,\be-1)}_{\frac{m-1}2}-P^{(\al+3,\be-1)}_{\frac{m-1}2}
+\frac13 P^{(\al+3,\be)}_{\frac{m-3}2}
+\frac{m+2\al+2}{m(m-2)} P^{(\al+1,\be)}_{\frac{m-3}2}\\
\phantom{\varphi_m=}{}
-\frac{m+2\al+2}{m-2} P^{(\al+2,\be)}_{\frac{m-3}2} ,\\
\phi_m=P^{(\al+4,\be-1)}_{\frac{m-3}2}-\frac{2m-5}{m-2} P^{(\al+3,\be)}_{\frac{m-3}2}-
\frac13 P^{(\al+4,\be)}_{\frac{m-5}2}+\frac{m+2\al-1}{m-2} P^{(\al+3,\be)}_{\frac{m-5}2} ,
\\
\chi_m=\frac{2m-3}{m(m-2)} P^{(\al+2,\be-1)}_{\frac{m-3}2}+\frac{2(m-3)}{m-2}
 P^{(\al+3,\be-1)}_{\frac{m-3}2}-P^{(\al+4,\be-1)}_{\frac{m-3}2}\\
\phantom{\chi_m=}{}-\frac{m+2\al+1}{m(m-2)} P^{
(\al+2,\be)}_{\frac{m-5}2}-\frac{m+2\al}{m-2} P^{(\al+3,\be)}_{\frac{m-5}2}
+\frac13 P^{(\al+4,\be)}_{\frac{m-3}2} ,
\end{gather*}
for odd $m$.

{\rm 2.} The Hamiltonian $H_1$ possesses the following families of
spin eigenfunctions with eigenvalue $E_k=E_0+4k\om$,
where $E_0=N\omega(4a+2b+1)$ is the ground state energy:
\begin{gather*}
\Psi^{(0)}_k=\mu L^{\al-1}_k(\omega r^2) \La\ket s ,\qquad k\geq0 ,\\
\Psi^{(1)}_k=\mu L^{\al+1}_{k-1}(\omega r^2)\big[N\La(x_1^2\ket
s)-r^2\La\ket s\big]
 ,\qquad k\geq1 ,\\
\Psi^{(2)}_k=\mu L^{\al+3}_{k-2}(\omega r^2)\Big[N(\al+1)\sum_i
x_i^4-\be\ms r^4\Big] \La\ket s , \qquad k\geq 2 ,
\end{gather*}
with $\al=N(2a+b+\frac12)$, $\be=N(4a+b+\frac32)$ and
$\mu=\e^{-\frac\omega2 r^2}\prod_i{|x_i^2-x_{i+1}^2|}^a x_i^b$.

{\rm 3.} The Hamiltonian $H_2$ possesses the following spin
eigenfunctions with zero momentum
\begin{gather*}
\Psi_0=\mu \Phi^{(0)},\qquad
\Psi_{1,2}=\mu\sum_i\bigg\{\begin{matrix}\cos\\\sin\end{matrix}\bigg\}
\big(2(x_i-\Bx)\big)\ket{s_i},\\
\Psi_3=\mu\bigg[\frac{2a}{2a+1} \Phi^{(0)}+\sum_{i\ne
j}\cos\!\big(2(x_i-x_j)\big)\ket{s_j}\bigg],
\qquad\Psi_4=\mu\sum_{i\ne j}\sin\!\big(2(x_i-x_j)\big)\ket{s_j},
\end{gather*}
where $\mu=\prod_i\sin^a|x_i-x_{i+1}|$. Their energies are
respectively given by
\[
E_0 ,\qquad E_{1,2}=E_0+4\Big(2a+1-\frac1N\Big) ,\qquad
E_{3,4}=E_0+8(2a+1) ,
\]
where $E_0=2Na^2$ is the ground state energy.
\end{theorem}

\section{Calogero models and Dunkl operators} \label{S:Calogero}

Let us consider a self-adjoint operator $H$ acting on a given
Hilbert space $\cH$. The idea underlying the construction of
algebraic eigenfunctions of $H$ (see, e.g.,~\cite{Tu94}) is that the explicit knowledge of a
f\/inite-dimensional subspace $\cH_1\subset\cH$ which is invariant
under the operator $H$ allow ones to compute $\dim\cH_1$
eigenfunctions and eigenvalues of $H$ by algebraic methods, i.e.,
diagonalizing the matrix $H|_{\cH_1}$. In this case the operator is
said to be {\em quasi-exactly solvable} (QES). In fact, the
Hamiltonians of the models~\eqref{V0} and~\eqref{V1} possess an
inf\/inite f\/lag $\cH_1\subset\cH_2\subset\cdots$ of known
f\/inite-dimensional invariant subspaces, which yield an arbitrary
large number of eigenvalues and eigenfunctions. Although such models
are sometimes termed ``exactly solvable''~\cite{Tu92}, we will not
use this terminology since the algebraic eigenfunctions of the
models~\eqref{V0} and~\eqref{V1} are not an orthonormal basis of the
Hilbert space.

A particularly convenient method of carrying out this
program~\cite{Tu88,ST89} is through a Lie algebra $\mathfrak g$ of
f\/irst-order dif\/ferential operators
$J^a=\sum_i\xi^{ai}(\bz)\ms\pd_{z_i}+\eta^a(\bz)$ ($a=1,\dots, r$)
acting on a f\/inite-dimensional module $\cM\subset C^\infty(M)$,
with $M$ a domain in $\RR^N$. Let us assume that there exists a
second-order dif\/ferential operator $\tilde
H=\sum\limits_{a,b=1}^rc_{ab}J^aJ^b+\sum\limits_{a=1}^rc_aJ^a+c_0$ that is
equivalent, up to gauge transformation $\tilde H\mapsto \mu\tilde
H\mu^{-1}$ ($\mu\in C^\infty(M,\RR^+)$) and global change of
variables $\bz\in M\mapsto \bx\in\RR^N$, to a Schr\"odinger
operator $H=-\Delta_g+V$, $\Delta_g$ standing for the
Laplace--Beltrami operator in $(\RR^N,g)$. If
$\mu\cM|_{\bz\mapsto\bx}\subset L^2(\RR^N,\sqrt g \dd\bx)$, then
one can obtain $\dim\cM$ eigenfunctions of $H$ by algebraic
methods. The Schr\"odinger operators amenable to this treatment
are termed \emph{Lie-algebraic}. It should be observed that the
operators $J^a$ are not symmetries of $H$: $\mathfrak g$ is an
algebra of \emph{hidden symmetries} of $H$.

Dunkl operators~\cite{He97} were originally introduced to study
spherical harmonics associated to measures invariant under a
Coxeter group~\cite{Du89}. Actually, let $\bv,\bz\in\RR^N$ and
def\/ine the ref\/lection
$\si_\bv\bz=\bz-2|\bv|^{-2}(\bz\cdot\bv)\bv$. Let $\fW$ be a
Coxeter group, which can be assumed to be the Weyl group of a
(possibly nonreduced) root system $R$, and let
$\bv_a=\sum_iv_a^i{\boldsymbol e}_i$ ($a=1,\dots,r$) be a basis of
positive roots. Def\/ine an action of $\fW$ on $\CC[\bz]$ as
$K_af=f\circ\si_{\bv_a}$. Dunkl operators were originally def\/ined
as
\begin{equation}\label{Dunkl}
J_i=\pd_{z_i}+\sum_{a=1}^r\frac{g_a^2v_a^i}{\bz\cdot\bv_a}(1-K_a) ,
\end{equation}
where the real parameters $g^2_a$ are chosen so that they are
constant on each orbit of $\fW$. The operators $J_i$ can be
understood as deformations of the partial derivatives $\pd_{z_i}$
that commute with the deformed Laplacian, i.e.,
$\sum_iJ_i^2J_j=J_j\sum_iJ_i^2$. It can be verif\/ied that
$\{J_i,K_a\}$ span a degenerate Hecke algebra~\cite{Ki97}.
Currently, the def\/inition of Dunkl operators has been generalized
to mean a set $\{J_i\}_{i=1}^N\subset\End\CC[\bz]$ of f\/irst order
dif\/ferential operators which leave invariant f\/inite-dimensional
polynomial subspaces and such that $\{J_i,K_a\}$ span a Hecke
algebra.

In their original form, Dunkl operators are directly connected
with the rational Calogero model of type $R$~\cite{Du98}. In the
simplest case ($R=A_{N-1}$, $\fW=S_N$) and writing $\bx$ instead
of $\bz$, the Dunkl operators~\eqref{Dunkl} read
\[
J_i=\pd_{x_i}+g^2\sum_{j\neq i}\frac1{x_i-x_j}(1-K_{ij}) ,
\]
where
\begin{equation}\label{Kij}
(K_{ij}f)(x_1,\dots,x_i,\dots,x_j,\dots,x_N)=f(x_1,\dots,x_j,\dots,x_i,\dots,x_N)
\end{equation}
denotes the ref\/lection operator associated to the root ${\boldsymbol 
e}_i-{\boldsymbol e}_j$. Actually, let us consider the Calogero ground
state function $\mu(\bx)=\e^{-\om r^2}\prod_{i<j}|x_i-x_j|^a$ and
the auxiliary operator
\begin{equation}\label{J0}
J^0=\sum_ix_i\pd_{x_i} .
\end{equation}
A straightforward calculation shows that one can use the ground
state function $\mu$ to gauge transform the deformed Laplacian so
that
\begin{equation}\label{HK}
\mu\BH_{\mathrm C}\mu^{-1}=-\sum_i\pd_{x_i}^2+a\sum_{i\neq
j}\frac1{(x_i-x_j)^2}(a-K_{ij})+\om r^2 ,
\end{equation}
with $g^2=a(a-1)$ and
\[
\BH_{\mathrm C}=\sum_iJ_i^2+2\om J^0+E_0 ,
\]
yields the Calogero model of $A_{N-1}$ type when acting on
symmetric functions $\psi\in\La L^2(\RR^N)$.

Since $J_i$, $J^0$ preserve the space of polynomials
\begin{equation}\label{Pn}
\cP^n=\{f\in\CC[\bx]:\deg f\leq n\}
\end{equation}
for any $n=0,1,\dots$, the Dunkl operators provide a very
convenient fashion of exploring the solvability properties of
Calogero--Sutherland models. It should be remarked, however, that
for an arbitrary Coxeter group $\fW$ the Dunkl
operators~\eqref{Dunkl} do not form a Lie algebra; nevertheless,
this technique captures most of the relevant features of the
Lie-algebraic method.

As we shall now outline, Dunkl operators can also be used to
introduce internal degrees of freedom in the picture without
breaking the solvability properties of the models. Given a scalar
dif\/ferential-dif\/ference operator $D$ linear in $K_{ij}$, let us
denote by $D^*$ the dif\/ferential operator acting on
$C^{\infty}\otimes\Si$ obtained from $D$ by the replacement
$K_{ij}\to S_{ij}$. It is clear that the actions of $D$ and $D^*$
coincide on the (bosonic) Hilbert space
\begin{equation}\label{cH}
\cH=\La(L^2(\RR^N)\otimes\Si) .
\end{equation}
Therefore the scalar operator~\eqref{HK} coincides with that of
the spin Calogero model
\[
H_{\mathrm C}\equiv \mu \BH_{\mathrm
C}^*\mu^{-1}=-\sum_i\pd_{x_i}^2+a\sum_{i\neq
j}\frac1{(x_i-x_j)^2}(a-S_{ij})
\]
when acting on symmetric states $\Psi\in\cH$.

\section{Invariant subspaces}\label{S:Inv}

The proof of the main theorem rests on the construction of
appropriate invariant f\/lags for the Hamiltonians~\eqref{Hep} using
a modif\/ication of the Dunkl operator formalism. This modif\/ication turns
out to be rather nontrivial, ultimately due to the fact the cyclic
group is not of Coxeter type.

As the f\/irst step, we consider the second-order
dif\/ferential-dif\/ference operators $T_\ep$ given by
\begin{equation}\label{Tep}
T_\ep=\sum_iz_i^\ep\pa_i^2+2a\sum_i\frac1{z_i-z_{i+1}} (z_i^\ep\pa_i-z_{i+1}^\ep\pa_{i+1})
-2a\sum_i\frac{\vartheta_\ep(z_i,z_{i+1})}{(z_i-z_{i+1})^2} (1-K_{i,i+1}),
\end{equation}
where $\pa_i=\pa_{z_i}$, $z_{N+1}\equiv z_1$, and
\[
\vartheta_0(x,y)=1 ,\qquad
\vartheta_1(x,y)=\frac12 (x+y) ,\qquad \vartheta_2(x,y)=xy .
\]
Each Hamiltonian $H_\ep$ is related to a linear combination
\begin{equation}\label{BHep}
\BH_{\!\ep}=c T_\ep+c_- J^-+c_0 J^0+E_0
\end{equation}
of its corresponding operator $T_\ep$ and the auxiliary
f\/irst-order dif\/ferential operators $J^-=\sum_i \pa_i$ and
$J^0=\sum_iz_i\pa_i$ via a change of variables, a gauge transformation,
and the star mapping def\/ined in the previous section, that is,
\begin{equation}\label{Hepstar}
H_\ep=\mu\cdot\BH_{\!\ep}^*\big|_{z_i=\ze(x_i)}\cdot\mu^{-1} .
\end{equation}
The constants $c$, $c_-$, $c_0$, $E_0$, the gauge factor
$\mu$, and the change of variables $\ze$ for each model are listed
in Table~\ref{table:params}. Hence the construction of the
models~\eqref{Hep} is analogous to that of the usual
Calogero--Sutherland models, with the operators~\eqref{Tep} being
a cyclic analog of the sum of the squares of the Dunkl operators.

\begin{table}[h]
\vspace{-2mm}
\caption{Parameters, gauge factor and change of variable in
equations~\eqref{BHep} and~\eqref{Hepstar}.}\label{table:params}
\begin{center}
\vspace{-1mm}
\begin{tabular}{lccc}\hline
\BStrut & $\ep=0$ & $\ep=1$ & $\ep=2$\\ \hline
\BStrut $c$ & $-1$ & $-4$ & $4$\\ \hline
\BStrut $c_-$ & $0$ & $-2(2b+1)$ & $0$\\ \hline
\BStrut $c_0$ & $2\om$ & $4\om$ & $4(1-2a)$\\ \hline
\BStrut $E_0$ & $N\omega(2a+1)$ & $N\omega(4a+2b+1)$ & $2Na^2$\\
\hline
\BStrut $\mu(\bx)\quad$ &
$\;\e^{-\frac\omega2 r^2}\prod\limits_i|x_i-x_{i+1}|^a\;$ &
$\;\e^{-\frac\omega2 r^2}\prod\limits_i{|x_i^2-x_{i+1}^2|}^a x_i^b\;$
& $\;\prod\limits_i\sin^a|x_i-x_{i+1}|\;$\\ \hline
\BStrut $\ze(x)$ & $x$ & $x^2$ & $\e^{\pm2\iu x}$\\ \hline
\end{tabular}
\end{center}
\vspace{-4mm}
\end{table}

It is obvious that the operators~\eqref{BHep} preserve the
polynomial space~\eqref{Pn} for any $n\in\NN$, so one may be
tempted to believe that the usual arguments for spin CS models
should yield an invariant f\/lag for the Hamiltonians~\eqref{Hep}.
Nevertheless, this is not the case. In fact, the standard
construction~\cite{FGGRZ01,FGGRZ01b} is based on the fact that the
actions of $H_{\mathrm C}$ and $\mu\BH_{\mathrm C}\mu^{-1}$ on
symmetric states coincide and these operators commute with the $S_N$
symmetrizer $\La$. Unfortunately, $H_\ep$ (or~$\BH_\ep$) do not
commute with $\La$ and cyclic symmetry does not suf\/f\/ice to exchange
$H_\ep$ and $\mu \BH_\ep\mu^{-1}$, so this procedure does not grant
the existence of any nontrivial invariant subspaces, not even of
direct products $\cM\otimes\La\Si$ ($\cM\subset C^\infty(\RR^N)$).

We shall now review the actual construction of the invariant
spaces, which is considerably more involved. We shall not provide
complete proofs, but merely a sketch of the main logical steps the
construction rests upon.

Classical results on the theory of invariants~\cite{We97} are
responsible for the success of studying CS models through symmetric
polynomials~\cite{Pe71,Tu95}. We shall extend this approach to deal
with the Hamiltonians~\eqref{Hep}. Let us f\/irst introduce two bases
$\{\si_k\}$ and $\{\tau_k\}$ of the space of symmetric polynomials
in $\bz$:
\[
\si_k=\sum_i z_i^k ,\qquad
\tau_k=\sum\limits_{i_1<\cdots<i_k}z_{i_1}\cdots z_{i_k} ; \qquad
k=1,\dots,N .
\]
The operators $T_\ep$ consist of three summands which are of second,
f\/irst and zeroth order in the derivatives. Let us denote each
summand by $L_\ep$, $2aX_\ep$ and $-2aA_\ep$ respectively. It is not
dif\/f\/icult to realize that $\{L_\ep\}$ span a Lie algebra isomorphic
to $\mathfrak{sl}(2)$, as in the CS case. It turns out that the
f\/irst-order dif\/ferential operators $X_\ep$ leave invariant a f\/lag of
symmetric polynomials, as stated in the following easy lemma.

\begin{lemma}\label{Lemma1}
For each $n=0,1,\dots$, the operator $X_\ep$ leaves invariant the
linear space $\cX_\ep^n$, where
\[
\cX_0^n=\CC[\si_1,\si_2,\si_3]\cap\cP^n,\quad
\cX_1^n=\CC[\si_1,\si_2,\tau_N]\cap\cP^n,\quad
\cX_2^n=\CC[\si_1,\tau_{N-1},\tau_N]\cap\cP^n.
\]
\end{lemma}
\begin{remark}
It should be noted that these f\/lags cannot be trivially enlarged,
since, e.g.,
\begin{alignat*}{3}
&\frac14 X_0\si_4=2\si_2+\sum_iz_iz_{i+1} ,&&&\\
& \frac13 X_1\si_3=2\si_2+\sum_iz_iz_{i+1} ,&&
X_1\tau_{N-1}=\tau_N\sum_i({z_iz_{i+1}})^{-1} ,&\\
&\frac12 X_2\si_2=2\si_2+\sum_iz_iz_{i+1} ,\qquad &&
X_2\tau_{N-2}=N\tau_{N-2}-\tau_N\sum_i(z_iz_{i+1})^{-1}&
\end{alignat*}
are not symmetric polynomials.
\end{remark}

In the next proposition we characterize subspaces of the f\/lags
described in Lemma~\ref{Lemma1} that are preserved by the whole
operator $T_\ep$. If
$f\in\CC[\si_1,\si_2,\si_3,\tau_{N-1},\tau_N]$, we adopt the
convenient notation
\[
f_k=\begin{cases}
\pa_{\si_k}f ,\quad & k=1,2,3,\\
\pa_{\tau_k}f ,\quad & k=N-1,N.
\end{cases}
\]

\begin{proposition}\label{prop.cS}
For each $n=0,1,\dots$, the operator $T_\ep$ leaves invariant the
linear space $\cS_\ep^n$, where
\begin{gather*}
\cS_0^n=\{f\in\cX_0^n\mid f_{33}=0\} ,\\
\cS_1^n=\{f\in\cX_1^n\mid f_{22}=f_{NN}=0\} ,\\
\cS_2^n=\{f\in\cX_2^n\mid f_{11}=f_{N-1,N-1}=0\} .
\end{gather*}
\end{proposition}

Proposition~\ref{prop.cS} implies that each operator $T_\ep$
preserves product symmetric subspaces $\cS^n_\ep\otimes\La\Si$
spanned by factorized states. The main result on invariant
subspaces shows that in fact the latter operator leaves invariant
a richer f\/lag of nontrivial f\/inite-dimensional subspaces of
$\La(\cP^n\otimes\Si)$.

\begin{theorem}\label{thm.1}
Let
\begin{gather*}
\cT^n_0 =\big\langle f(\si_1,\si_2,\si_3)\La\ket
s,g(\si_1,\si_2,\si_3)\La(z_1\ket s),
h(\si_1,\si_2)\La(z_1^2\ket s),\\
\phantom{\cT^n_0 =\big\langle}{}\tilde h(\si_1,\si_2)\La(z_1z_2\ket{s'})\mid f_{33}=g_{33}=0\big\rangle ,\\
\cT^n_1 =\big\langle f(\si_1,\si_2,\tau_N)\La\ket s,
g(\si_1,\tau_N)\La(z_1\ket s)
\mid{}f_{22}=f_{NN}=g_{NN}=0\big\rangle ,\\
\cT^n_2=\big\langle f(\si_1,\tau_{N-1},\tau_N)\La\ket
s,g(\tau_{N-1},\tau_N)\La(z_1\ket s),
\tau_Nq(\si_1,\tau_N)\La(z_1^{-1}\ket s)\\
\phantom{\cT^n_2=\big\langle}{}\mid
f_{11}=f_{N-1,N-1}=g_{N-1,N-1}=q_{11}=0\big\rangle ,
\end{gather*}
where $\ket s\in\Si$, $\ket{s'}\in\Si'$, $\deg f\leq n$, $\deg
g\leq n-1$, $\deg h\leq n-2$, $\deg\tilde h\leq n-2$, $\deg q\leq
n-N+1$, and $\deg$ is the total degree in $\bz$. Then $\cT_\ep^n$
is invariant under $T_\ep$ for all $n=0,1,\dots$.
\end{theorem}

From this theorem one easily obtains the following corollary,
which is crucial for the computation of the algebraic
eigenfunctions of the models~\eqref{Hep}.

\begin{corollary}\label{cor.1}
For each $\ep=0,1,2$, the gauge Hamiltonian $\BH_\ep$ leaves
invariant the space $\BcH^n_\ep$ defined by
\begin{equation}\label{BcHs}
\BcH^n_0=\cT^n_0 ,\qquad
\BcH^n_1=\cT^n_1\big|_{f_N=g_N=0} ,\qquad \BcH^n_2=\cT^n_2 .
\end{equation}
\end{corollary}

\section{Spectrum and eigenfunctions}\label{S:spec}

As happens with the usual CS models, the algebraic eigenvalues of
the Hamiltonians~\eqref{Hep} can be easily obtained by choosing a
basis of the invariant spaces $\mu\BcH^n_\ep$ in which the action
of $H_\ep$ is triangular. In all three cases, the algebraic
eigenvalue $E_0$ is the ground state energy, since the
corresponding eigenfunctions do not vanish in the conf\/iguration
space $C_\ep$.

Now we shall outline how the algebraic eigenfunctions in
Theorem~\ref{main} were calculated. The easiest case is $\ep=2$,
since states related by a multiplicative factor $\tau_N^k$ only
dif\/fer by the movement of the center of mass, which is conserved.
Hence the invariant subspaces
$\BcH^n_2$ solely allow one to compute f\/ive eigenfunctions of zero
total momentum, which are the ones listed in the main theorem.

In the case $\ep=1$ the eigenvalue equation reads
\[
(\BH_1-E_0-4\om n)\Phi=0 ,
\]
where $\Phi$ is a symmetric vector-valued polynomial in $\BcH^n_1$
of degree $n$. Setting $t=\om\si_1$ and
$\Phi=[p(t)+\si_2q(t)]\La\ket s+g(t)\La(z_1\ket s)$, this equation
can be written as
\begin{equation}\label{Eqs.2}
\cL^{\al+1}_{k-1}g=\cL^{\al+3}_{k-2}q=0 ,\qquad
\cL^{\al-1}_kp=-\frac\al{N\om} g-\frac{2\be}{N\om^2} tq ,
\end{equation}
with the Laguerre operator $\cL^\la_\nu$ def\/ined as
\[
(\cL^\la_\nu f)(t)=t\ms f''(t)+(\la+1-t)f'(t)+\nu f(t) .
\]
With some ef\/fort one can obtain all the solutions of these
equations, as shown in the following proposition.

\begin{proposition}
The polynomial solutions of the system of ODE's~\eqref{Eqs.2} are
spanned by
\begin{gather*}
\Phi^{(0)}_n=L^{\al-1}_n(t) \La\ket s ,\qquad n\geq0 ,\\
\Phi^{(1)}_n=L^{\al+1}_{n-1}(t)\big[N\om\La(z_1\ket
s)-t\La(z_1\ket s)\big]
 ,\qquad n\geq1 ,\\
\Phi^{(2)}_n=L^{\al+3}_{n-2}(t)\Big[N\om^2(\al+1)\si_2-\be\ms
t^2\Big] \La\ket s , \qquad n\geq 2 .
\end{gather*}
\end{proposition}

These solutions are characterized by the conditions $q=g=0$, $q=0$
and $g=0$ respectively and correspond to the algebraic
eigenfunctions of $H_1$ presented in Theorem~\ref{main}.

The case $\ep=0$ is similar, but the computations become more
involved due to the rich structure of the invariant f\/lag. Writing
\[
\Phi=(p+\si_3 q)\La\ket s+(u+\si_3 v)\La(z_1\ket s)+h\La(z_1^2\ket
s)+\Th\La(x_1x_2\ket s) ,
\]
where $\deg\Phi=k$ and $\Th=0$ if $\ket s\not\in\Si'$, the
equation $(H_0-E_0-2\om k)\Phi=0$ reduces to the systems of PDE's
\begin{subequations}\label{system2}
\begin{gather}
 \big[L_0-2\om(k-2)\big]\Th-8\Th_2=0 ,\label{system2th}\\
 \big[L_0-2\om(k-2)\big]h-8h_2=6v ,\label{system2h}\\
 \big[L_0-2\om(k-1)\big]u-4u_2=4h_1+4\Th_1+6\si_2v_1+6(2a+1)\si_1 v ,\label{system2u}\\
 \big[L_0-2\om(k-4)\big]v-16v_2=0 ,\label{system2v}\\
 \big(L_0-2\om k\big)p=2u_1+2(2a+1)h-\frac{4a}{N-1} \Th+6\si_2q_1+6(2a+1)\si_1q ,\label{system2p}\\
 \big[L_0-2\om(k-3)\big]q-12q_2=2v_1 ,\label{system2q}
\end{gather}
\end{subequations}
with
\[
L_0=-\big(N\pa_{\si_1}^2+4\si_1\pa_{\si_1}\pa_{\si_2}+4\si_2\pa_{\si_2}^2+2(2a+1)N\pa_{\si_2}\big)
+2\om(\si_1\pa_{\si_1}+2\si_2\pa_{\si_2}) .
\]
One can check by inspection that the system~\eqref{system2} possesses six families of
polynomial solutions, as collected in Table~\ref{table:sols0}.
These correspond to the six families of algebraic eigenfunctions
listed in Theorem~\ref{main}.
\begin{table}[h]
\vspace{-2mm}

\caption{The six types of polynomial solutions of the
system~\eqref{system2} and their corresponding
eigenfunctions.}\label{table:sols0}
\begin{center}
\vspace{-1mm}
\begin{tabular}{lc}\hline
\vrule height 14pt depth 6pt width0pt {\hfill
Conditions\hphantom{\qquad}\hfill} & Corresponding eigenfunction\\
\hline
\BStrut $q=u=v=h=\Th=0,\quad p\neq0$\hphantom{\qquad} &
$\Psi^{(0)}_{lm}$\\ \hline
\BStrut $u=v=h=\Th=0,\quad q\neq0$\hphantom{\qquad} &
$\Psi^{(3)}_{lm}$\\ \hline
\BStrut $q=v=h=\Th=0,\quad u\neq0$\hphantom{\qquad} &
$\Psi^{(1)}_{lm}$\\ \hline
\BStrut $q=v=\Th=0,\quad h\neq0$\hphantom{\qquad} &
$\Psi^{(2)}_{lm}$\\ \hline
\BStrut $q=v=h=0,\quad \Th\neq0$\hphantom{\qquad} &
$\widetilde\Psi^{(2)}_{lm}$\\ \hline
\BStrut $\Th=0,\quad v\neq0$\hphantom{\qquad} &
$\Psi^{(4)}_{lm}$\\ \hline
\end{tabular}
\end{center}
\vspace{-9mm}
\end{table}

\subsection*{Acknowledgements}

This work was partially supported by the DGI under grant
no.~FIS2005-00752. A.E. acknow\-led\-ges the f\/inancial support of the
Spanish Ministry of Education through an FPU scholarship.

\LastPageEnding

\end{document}